\documentclass[sigconf]{acmart}
\usepackage{enumitem}
\usepackage{balance}

\AtBeginDocument{%
  \providecommand\BibTeX{{%
    \normalfont B\kern-0.5em{\scshape i\kern-0.25em b}\kern-0.8em\TeX}}}

\copyrightyear{2023} 
\acmYear{2023} 
\setcopyright{rightsretained} 
\acmConference[CHIIR '23]{ACM SIGIR Conference on Human Information Interaction and Retrieval}{March 19--23, 2023}{Austin, TX, USA}
\acmBooktitle{ACM SIGIR Conference on Human Information Interaction and Retrieval (CHIIR '23), March 19--23, 2023, Austin, TX, USA}
\acmDOI{10.1145/3576840.3578288}
\acmISBN{979-8-4007-0035-4/23/03}

\newcommand{\circlenum}[1]{\raisebox{.9pt}{\textcircled{\raisebox{-.9pt}{#1}}}}

\begin{document}
\fancyhead{}

\title[Taking Search to Task]{Taking Search to Task}

\author{Chirag Shah}
\affiliation{
  \institution{University of Washington}
  \city{Seattle}
  \country{USA}
}
\email{chirags@uw.edu}

\author{Ryen W. White}
\affiliation{
  \institution{Microsoft Research}
  \city{Redmond}
  \country{USA}
}
\email{ryenw@microsoft.com}

\author{Paul Thomas}
\affiliation{
  \institution{Microsoft}
  \city{Canberra}
  \country{Australia}
}
\email{pathom@microsoft.com}

\author{Bhaskar Mitra}
\affiliation{
  \institution{Microsoft Research}
  \city{Montréal}
  \country{Canada}
}
\email{bmitra@microsoft.com}

\author{Shawon Sarkar}
\affiliation{
  \institution{University of Washington}
  \city{Seattle}
  \country{USA}
}
\email{ss288@uw.edu}

\author{Nicholas Belkin}
\affiliation{
  \institution{Rutgers University}
  \city{New Brunswick}
  \country{USA}
}
\email{belkin@comminfo.rutgers.edu}

\begin{abstract}
The importance of tasks in information retrieval (IR) has been long argued for, addressed in different ways, often ignored, and frequently revisited. For decades, scholars made a case for the role that a user's task plays in how and why that user engages in search and what a search system should do to assist. But for the most part, the IR community has been too focused on query processing and assuming a search task to be a collection of user queries, often ignoring if or how such an assumption addresses the users accomplishing their tasks. With emerging areas of conversational agents and proactive IR, understanding and addressing users' tasks has become more important than ever before. In this paper, we provide various perspectives on where the state-of-the-art is with regard to tasks in IR, what are some of the bottlenecks in deriving and using task information, and how do we go forward from here. In addition to covering relevant literature, the paper provides a synthesis of historical and current perspectives on understanding, extracting, and addressing task-focused search. To ground ongoing and future research in this area, we present a new framing device for tasks using a tree-like structure and various moves on that structure that allow different interpretations and applications. Presented as a combination of synthesis of ideas and past works, proposals for future research, and our perspectives on technical, social, and ethical considerations, this paper is meant to help revitalize the interest and future work in task-based IR.
\end{abstract}

\ccsdesc[500]{Information systems~Information retrieval}

\keywords{Tasks; Contextual search; Proactive search}
\maketitle

\section{Introduction}
Scholars have long argued for the importance of considering task information in information retrieval (IR) for truly helping people with complex, unexpressed, or unclear needs \cite{belkin1980anomalous, dervin1998sense}. Over the decades, the concept of task has been studied by many researchers who have produced notable theoretical and practical outcomes. Several attempts have been made to understand search tasks, characterize and extract them, and use task knowledge to better provide support in search and recommendation applications. There are several small and practical successes along the way, including search services incorporating spatial and temporal information in understanding or expanding a query, as well as using the current context and history activity to provide contextual recommendations. However, these efforts can be limiting at best and harmful at worst as they fail to regard user intents or goals as a way to model the ongoing task. 

Can we meaningfully connect operationalization of search to conceptualization of task (take search to task)? How do we create a framing device with tasks with the explicit purpose of applying it to various IR applications? What do we gain (and lose) if we are successful with this? These are some of the core questions that triggered our investigations -- some theoretical, some empirical, and others simply thought experiments -- resulting in this perspective paper. Thus, the purpose of this perspective paper is to shine the light, once again, on this very important area of IR and provide a new foundation built with current understanding and future possibilities that include emerging domains of conversational agents, multi-device search, and proactive recommenders to guide users to complete their tasks step by step.

The remainder of the paper is organized as follows. The next section reviews some of the most important and transformative research on tasks in IR over the last few decades. We also list several recent events and activities to demonstrate the importance of this area and emphasize the scholarly interest. In Section 3, we present a framing device to think through possibilities and challenges for capturing task-related information in IR. Section 4 extends this by providing paths and perspectives as we move forward, specifically focusing on task representation
%%\nick{I can't find any discussion of ethical issues in this section}
and using such representations in IR applications. Some of such applications that are taking shape now and are important in the future of IR 
are outlined in Section 5. %once again showcasing why and how task-based modeling is important. 
In Section 6, we briefly discuss methods and metrics for evaluating task-based applications. Finally, we conclude in Section 7 with our thoughts on task futures, along with a discussion of ethical considerations regarding using tasks in search and other applications.
\section{A Brief History of Tasks in IR}
%\chirag{Lit review around tasks in IR from 70s to 2000s. Assigned to SHAWON, CHIRAG. (1.5-2p.)}

A task is generally considered as a set of connected physical, cognitive, and affective  actions through which individuals try to accomplish some goals in their work or everyday lives~\cite{vakkari2003a, Bystrom2007}. In the context of IR, the concept of task has taken on explicit meanings related to understanding and supporting information seeking and searching. In this section, we give an overview of the ways in which task has been understood in previous IR-related research, beginning with a general survey of different approaches, then considering some specific aspects of task that have been investigated, followed by discussion of some significant attempts to apply knowledge of task in IR, and concluding with a discussion of recent workshops concerning task in IR.

%The way the notion of a task is understood has been instrumental in guiding the research in this area.

\vspace*{-0.5em}
\subsection{Overview of Approaches to Task in IR}
Some of the earliest prior research in IR related to task can be traced back to the cognitive perspective in IR \cite{belkin1990cognitive}, which was centrally concerned with understanding what motivated a person to engage in information seeking and searching. This perspective influenced works by Vakkari \cite{vakkari2001theory} and Ingwersen and J{\" a}rvelin \cite{ingwersen2006turn}, which consider tasks in the design of IR systems to find out for what purposes the system is used \cite{Saastamoinen2017} and thus provides implications for IR system design to personalize information search according to the task at hand. Based on a series of empirical works, Vakkari \cite{vakkari2001theory} developed a framework of task-based information searching comprising three stages: \textit{pre-focus}, \textit{focus formulation}, and \textit{post-focus}.

Tasks are often considered multi-level information seeking processes in which people need information to achieve a goal %to fulfill the task
\cite[e.g.,][]{vakkari1999task, bystrom1995task, Saastamoinen2017, savolainen2012expectancy}. Many existing task models \cite[e.g.,][]{kelly2015development, capra2018effects, li2008faceted} have investigated and identified searchers' tasks as static and overarching goals that motivate search actions, but this is not always desired as the task evolves with time and changing cognitive states. Conversely, different characteristics or facets of tasks \cite{li2008faceted} influence people's interaction with intelligent systems, such as search engines \cite{liu2019task}. Search tasks are influenced by the work task or everyday life task that drives them to seek information or are associated with a problematic situation \cite{bystrom2005conceptual}.

Identification of task, at various levels, has been an area of focus. Broder \cite{broder2002taxonomy} proposed that a person's intent or goal in engaging with a search engine could be one of three types: \emph{informational}, \emph{transactional}, and \emph{navigational}. %Although they are general categories of the types of tasks motivating engagement with a search system, they have 
This scheme has been successfully used in a great deal of research, to classify tasks according to search behaviors and for study and support of search according to type. \citet{rose2004understanding} extended Broder's scheme by specifying types of information goals, and adding a new goal type (\emph{resource}). They tested their scheme of motivating goals (or tasks), by classifying search engine queries.
%The aim of task-based studies is to investigate the relationships between task characteristics and information seeking behaviors by recognizing and understanding the nature of different tasks and goals and designing IR systems which can support the accomplishment of a variety of such tasks and goals. 

Many early works investigated and identified various aspects of task which could influence a variety of search behaviors, including task complexity \cite[e.g.,][]{bystrom1995task}, task difficulty \cite[e.g.,][]{kim2006difficulty} and work context \cite[e.g.,][]{Freund2005genre}. Others considered the interactive and dynamic nature of search tasks themselves \cite[e.g.,][]{bates1989design}.
%Also, accomplishing more complex tasks requires more complex actions that are manifested throughout the session because complex tasks take longer to complete or require more queries \cite{hassan2012task,awadallah2014a}.

Apart from task, existing studies in IR segment information seeking behaviors into various levels of explicit and implicit signals. While performing tasks, searchers' actions are also driven by intentions and can be well-defined or ill-defined \cite{ingwersen2006turn}. These studies have indicated that there is a close association between searchers' performance of a task and the information need, the search strategies employed, and the assessment of document relevance and utility.

Beyond search, tasks permeate almost every aspect of our daily work and personal lives \cite{allen2015getting}. They involve different activities, have different constraints, and take different amounts of time to complete. %Some tasks can be completed quickly, while others take much longer, sometimes spanning several days or weeks. Task management applications help people track their pending and completed tasks.
Users of task management applications would benefit from assistance with many aspects of task management, especially task planning \cite{bellotti2004studies} and prioritization \cite{myers2007intelligent}. There has been recent progress in task intelligence, in areas such as discovering digital assistant capabilities \cite{white2018skill}, estimating task durations \cite{white2019task}, and automatically tracking task status \cite{white2019task_2}.

It should be noted that not all research concerned with tasks in IR has been explicitly about modeling or using task. Some such examples include research on task trails \cite{liao2014task}, personalized search \cite{white2013enhancing}, trail recommendation \cite{singla2010studying}, cross-session tasks \cite{wang2013learning}, task continuation \cite{agichtein2012b}, and cross-device tasks \cite{wang2013characterizing}. %The following subsections provide further systematic review of relevant works.

\vspace*{-1em}
\subsection{Task Levels}
According to \citet{bystrom2002work, bystrom2005conceptual}, task contexts in information practices can be represented by a nested model consisting of three levels (from outer level to inner level): work task, information seeking task, and search task. Specifically, work tasks are separable parts of a person's duties in his or her workplace~\cite{bystrom2005conceptual}. 
%Note that not every sub-task within a work task can be transformed into an information seeking task. In many cases, some parts of a work task need system and human supports that are beyond the capacity of search systems (e.g., writing a dissertation proposal). 
%In addition to the tasks generated in workplaces, 
Everyday life tasks that emerge from non-work scenarios can also lead to active information seeking and searching practices (e.g., search for and book a hotel for travel)~\cite{agosto2006toward}.

In addition to Bystr{\"o}m and Hansen's nested model of task, \citet{xie2008interactive} also explored the multilevel nature of user goals and tasks and developed a four-level hierarchical framework of goals.
%including long-term goals (e.g., users' personal interests), leading search goals or work tasks, current search goals (current information seeking and search tasks), and interactive intentions (things that a user wants to accomplish in local steps or stages of search).
This four-level typology covers a wide range of user goals and tasks (from long-term task-independent goals to local goals behind specific search tactics) and was verified via user studies~\cite{xie2008interactive, lin2005validation}.

\vspace*{-1em}
\subsection{Task Facets}
Focusing on different dimensions or task taxonomies, previous research has examined the impacts of task \textit{types} and \textit{facets} on search interactions from different perspectives. %For instance, \citet{white2006study} found that knowledge of the motivating task type led to increased performance of implicit relevance feedback. 
\citet{liu2010search} and \citet{jiang2014searching} examined the associations between user behaviors and objective task features (i.e., task product, task goal, task type) and discussed to what extent these behavioral features can help disambiguate search tasks of different types. \citet{capra2018effects} found that manipulating task \textit{a priori} determinability via modifying task items and dimensions can significantly affect users' perceived task difficulty and choices of search strategies. 

With respect to task-user combined features, \citet{wildemuth2004effects} argued that in task-based information search,  search tactics are influenced by users' %domain knowledge related to task topics
topical knowledge. \citet{Liu:2010difficult} demonstrated that both whole-session level and within-session search behaviors are affected by task difficulty, and that the dynamic relationships between search behavior and task perception are influenced by task type. 
%(i.e., single fact-finding, multiple fact-finding, and multiple-piece information gathering). 
Similarly, \citet{Aula:2010} investigated search behavioral variations under tasks of different levels of difficulty, and found more query variance, more usage of advanced syntax, and longer time on search engine result pages (SERPs) with more difficult tasks. %By conducting a lab study and a large-scale online study, they found that when performing difficult search tasks, users tend to issue more diverse queries (have a more unsystematic query refinement process), use advanced operators more frequently, and spend longer time on search engine result pages (SERPs) during their search processes.

%Given that many IR studies only examine one or a few task dimensions, 
\citet{li2008faceted} developed a faceted approach to conceptualizing tasks in IR based on related literature on task classification as well as their empirical studies on task-based information searching~\cite{li2009exploring, Li2010}. The faceted framework provides a holistic approach to exploring the nature of tasks and conceptually supported a series of empirical studies on task-based search interactions. %Several recent works discussed in the current paper have used this framework to construct or examine tasks in interactive IR settings.

\vspace*{-1em}
\subsection{Task Stages}
Task process is an aspect of task which differs from static task properties or facets \cite{li2008faceted} (e.g., predefined task goal, task product). When conceptualizing tasks from the process-oriented perspective, we are essentially looking at the process of \textit{doing} or \textit{performing} tasks. The core argument here is that in the context of information seeking, we cannot define or study a task without examining \textit{how} the task was actually completed (or failed). Therefore, to fully understand a task, we need to explore both the objective task features and users' responses to the evolving task environments at multiple levels (e.g., behavioral, cognitive, emotional). 

%In the information seeking and IR communities, a series of classical models have been developed and applied to describe the general process of performing information seeking and search tasks. 
Many search process models focus on behavioral aspects and examine the transitions of information seeking and search actions. For instance, to describe the general process of information seeking, \citet{ellis1989behavioural} studied the information seeking patterns of academic social scientists and broke it down into six characteristics: starting, chaining, browsing, differentiating, monitoring, and extracting. \citet{wilson1999models} suggests that in some circumstances, Ellis' ``characteristics'' can be organized as a sequence of information seeking stages. Ellis' model clearly identifies the features of information seeking patterns and has been modified and tested empirically \cite[e.g.,][]{ellis1993modeling, ellis1993comparison}. However, this model only describes the behavioral level of task-based information seeking. It does not consider the interaction between the information seeker and the multi-dimensional context in which task states and information seeking activities evolve.

\vspace*{-1em}
\subsection{Applying Task Knowledge in IR}
%Some of the prominent ideas and outcomes in the area of applying
Applications of task knowledge to IR  have demonstrated that task representations can be used to provide users with better query suggestions \cite{awadallah2014a}, build user models for improved personalized search \cite{white2013enhancing, mehrotra2015b} and recommendation \cite{zhang2015task}, and help in satisfaction prediction \cite{hassan2013a, wang2014a}. %Perhaps the more frequent and widespread use of task representations is to build user models for personalized search and recommendation settings (e.g., \cite{white2013enhancing, mehrotra2015b}).
\citet{mehrotra2016b} used a tensor-based approach, representing each user as a combination of their topical interests and their search task behaviors for personalization. Other works have developed various novel task context embeddings to represent queries via search logs to provide task-based personalization, query suggestion, and re-ranking \cite{mehrotra2014task, mehrotra2015b}.  \citet{tolomei2010towards} investigated the concept of task flows and analyzed query logs to generate task-based query suggestions. \citet{baraglia2009a} introduced the notion of search shortcuts and offered query suggestions to drive goal attainment.

\citet{vu2015a} has also used tasks to model user interests in search. In a similar vein but in other contexts, several scholars have leveraged task information to provide long-term support for task completion \cite[e.g.,][]{jones2008beyond, white2019task, agichtein2012b}. \citet{cai2014personalized} used task models to improve the ranking of retrieved search results to provide task-based support to users. Tasks help users achieve their search goals and understand and evaluate a system's competency in helping users do so. \citet{hassan2013a} used search task constructs to predict satisfaction. \citet{white2006study} used them to improve relevance feedback. \citet{song2016a} demonstrated that task information could help to automate tasks to reduce user burden.

Other researchers have focused on assistive systems in terms of tours or trails to lead users through their search process \cite{o2010tweetmotif, hassan2012task, mitra2015exploring}, predicting users' next search action based on the current actions, either by predicting the next result click \cite{cao2009b} or by predicting short-term interests based on task topic information \cite{white2010predicting}.

\vspace*{-0.5em}
\subsection{Recent Research, Development, Activities}
%\chirag{Some of the recent works, including things from Rutgers (Nick, Chirag), UCL (Emine, Rishabh), TREC task track, etc. Assigned to NICK, RYEN. (0.5p.)}
%\ryen{Also work on task trails \cite{liao2014task}, personalized search \cite{white2013enhancing}, trail recommendation \cite{singla2010studying}, cross-session tasks \cite{wang2013learning}, task continuation \cite{agichtein2012search}, cross-device tasks \cite{wang2013characterizing}, etc.}

There continues to be significant interest and activity surrounding tasks from the research community. Several workshops have been held on task-based IR, focusing on search interactions, searcher intents, and tasks in information search. This includes the SIGCHI 2012 workshop on \textit{End-user Interactions with Intelligent Systems} \cite{stumpf2012end}, and the \textit{Second Strategic Workshop on Information Retrieval in Lorne (SWIRL)} \cite{allan2012frontiers}. The \textit{Task-based and Aggregated Search} workshop held in 2012 \cite{larsen2012report} focused on the challenges of task-based and aggregated search, such as the mismatch between search interface and specialized task-based functionalities, the lack of homogeneous systems to support different tasks, and so on. 
%One of the significant contributions of this early workshop was that it identified how and to what extent domain-specific search and recommendation systems could be developed to support task level activities. Participants also discussed how a search system should be modified in order to provide better support for task-based search. 
In the same year, the SIGIR 2012 workshop entitled \textit{``Entertain Me'' Supporting Complex Search Tasks} \cite{belkin2012report} focused on fostering potential solutions to problems faced by searchers with complex information needs. 
%Aiming to support searchers during their entire search sessions when interactively solving a complex task, the workshop explored many aspects of interactive information systems such as complex search episodes, queries, exploratory search, understanding of search context, and finally, how to incorporate task and searcher context into an information system.

An NSF-sponsored workshop on \textit{Task-Based Information Search Systems}, held in 2013, discussed challenges in developing systems and tools to support tasks and user needs \citet{kellysigirforum}.
%invited 29 "... leading international researchers in information retrieval, human-computer interaction and information behavior to discuss research and challenges in incorporating models of tasks, task-types, and users' needs into systems/tools to support complex, multi-search and multi-session tasks." (\cite{kellynsf}, p. 5).
%The workshop defined a research agenda for task-based search, and of challenges to such an agenda. Specific areas identified for further research included: Development of domain neutral modeling techniques to represent tasks and task-related behaviors; creation of task-specific and task-aware search environments; and, methods and measures for evaluating task-based systems. Development of task models was seen as the most crucial of these, as this was understood as necessary for accomplishment of the others. These research aims, and the challenges to them identified by the participants, are still germane, and resonate strongly with what is proposed in this paper.
The SIGIR 2013 workshop on \textit{Modeling User Behavior for Information Retrieval Evaluation} \cite{clarke2013report}, examined ways to model search intent based on queries. 
%Participants also identified problems with the use of queries as a proxy for search intent and brainstormed better solutions. 
Workshops on \textit{Supporting Complex Search Tasks} held in 2015 \cite{gade2015report} and 2017 \cite{belkin2017second} initiated interdisciplinary dialog on many task-related open research questions, including evaluation and the role of context.
%Participants tackled issues related to six aspects of information seeking – context, tasks, heterogeneous sources and search process, user interfaces (UI) and user experience (UX), and evaluation of systems. The workshops were helpful in fostering new collaborations among different communities to address these issues.
The WSDM 2018 workshop on \textit{Learning from User Interactions} \cite{mehrotra2018learnir}, focused on task-based intelligent systems, more specifically on six related topics – user needs and task understanding, user modeling and personalization, metrics and evaluation, user interaction processes and context, intelligent interface design and applications. %The workshop attracted participants from IR, human factors, ubiquitous computing, data mining, and other related domains. 
The WSDM 2019 workshop on \textit{Task Intelligence} \cite{hassan2019task}, focused on tasks in the context of system development, including areas such search assistance, personalization, and recommendation. Shah and White \cite{shah2020tutorial} also delivered a well-attended tutorial on this topic at SIGIR 2020.

\section{Task Composition and Support}
%\paul{let's swap this with the next \S, since from here (``here are some problems'') it's natural to move to ``here are our ideas''}
%\chirag{Assigned to PAUL, BHASKAR (1p.)}

%\chirag{Expand this core section with more details about our perspective. For this, add some description about some of our efforts and thinking with earlier experiments and failures. This should motivate what follows.}

These decades of work have led to many different mechanisms for representing tasks, which we can divide into two sets: explicit and implicit representations. Explicitly represented tasks are often presented as hierarchies, trees, or lists of aspects. These are explainable and readily interpretable. Implicit representations often use a probability distribution (over latent aspects of tasks) or encoded vectors. Such representations are usually not meant to be interpreted by humans, but they can offer more flexibility.

%It should be clear from the previous section that {\em task in IR} has been studied extensively over several decades, with many different frameworks and models  proposed for representing the tasks. We can divide them in two broad categories: explicit representation and implicit representation. Explicitly represented tasks are often presented as hierarchies, trees, or a list of dimensions/aspects. Such representations have the benefits of being explainable and more readily interpretable. Implicit representation of task is often done through a probability distribution (over latent aspects of tasks) or encoded vectors. Such representations are usually not meant to be interpreted by humans, but they often offer a more flexible representation of tasks with scalable applications.

We have experimented with both of these representations over the years \cite[e.g.,][]{cole2014discrimination, Mitsui2018,liu2019task}, recognizing their advantages and disadvantages. However, %in our most recent attempts,
we have started to converge on ideas that offer the best of both worlds---providing the interpretability of an explicit representation, with the scalability of an implicit representation. For example, we focused on task completion, defining three stages of a task: \textit{beginning}, \textit{continuing/exploration}, and \textit{ending/terminal}. For a given task and its stage, we also attempted to identify the kinds of support the user could use. Such support may include query or document suggestions, snippets or answers, as well as external tools. The goal here was to do manual (explicit) annotations for many search sessions with known tasks to then learn a model that could create an implicit representation (e.g., vector embeddings) of a task with respect to some application, such as next query prediction.

However, as we worked with several real-world datasets of search sessions, we realized that our coding scheme for task stage and support annotations was not as comprehensive or robust as we had hoped. We need a better framework that offers both comprehensive representation of a task as well as enough flexibility to be able to accommodate various applications and datasets. We discuss a possible approach next.

%\chirag{Need to put some transition from the above paragraphs to the next}
%\nick{I added a kind of transitional sentence to the last paragraph, and changed the intial words of the next paragraph.}

\vspace*{-0.5em}

\subsection{Tasks as Trees}
\begin{figure}
    \centering
    \includegraphics[width=0.9\columnwidth]{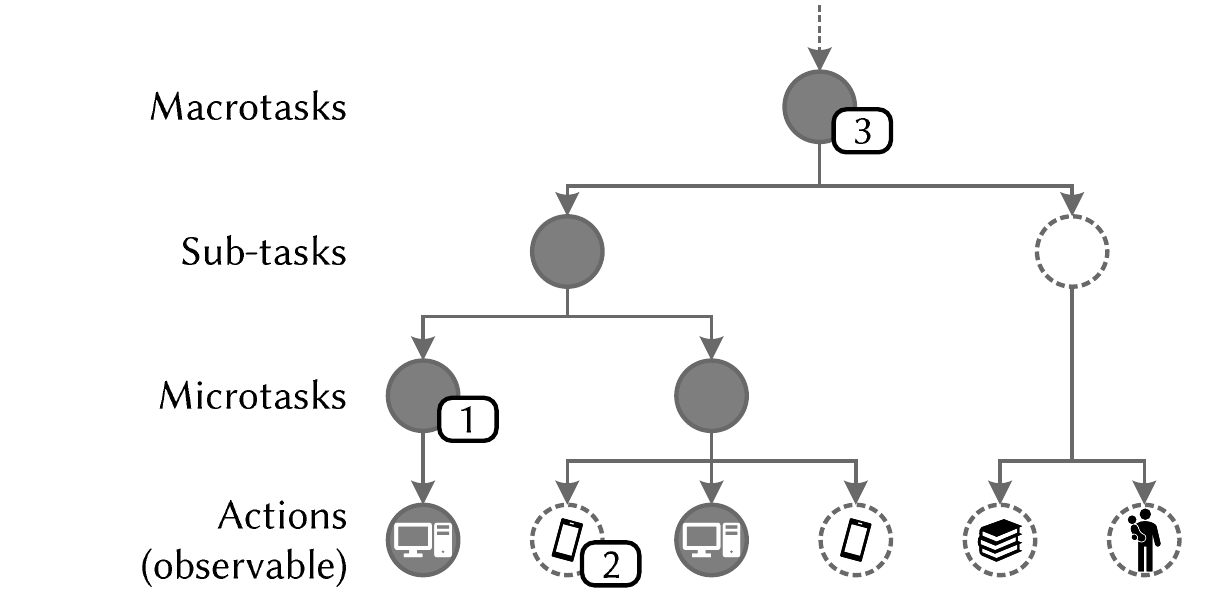}
    \caption{An abstract task ``tree''. Larger tasks may be decomposed into %one or more
    smaller tasks, and ultimately to actions. Some of these may be unobservable (dotted lines). Task %representation and
    support needs us to move ``up'', ``down'', and ``across'' the tree. See text for notes \circlenum{1}, \circlenum{2}, and \circlenum{3}.
    \vspace*{-1em}}
    \label{fig:task_potato}
\end{figure}

We now consider some mechanisms and times for a search system to support a searcher's tasks. Simplifying the nested model of \citet{bystrom1995task} and the hierarchies of \citet{xie2008interactive}, we can say that a task (also called a ``macrotask'' \cite{cheng2015break}) is composed of sub-tasks, sub-sub-tasks, and so on. For example, ``arrange a vacation in Austin'' may consist of ``find the best dates'' and ``make bookings''; ``make bookings'' might be composed of ``book flights'' and ``book a hotel''; etc. Each of these sub-tasks could be at any of \citeauthor{bystrom1995task}'s levels (Figure~\ref{fig:task_potato}). At the lowest level, a simple task is instead composed by ``actions'': the observable things people (or as we will discuss later, systems) might do. These could be instances of queries, or clicks; but could also be reading books, conversing with friends, or other moves (bottom level of Figure~\ref{fig:task_potato}). In some cases this structure will be explicit, as in a project plan or a hierarchical to-do list, but more often it will not be. The structure might not be mapped out at the start, will certainly be dynamic in all but the simplest cases, and different strategies will be useful at different points. As searchers may be simultaneously engaged in multiple tasks, the corresponding hierarchies of sub-tasks and actions may also interleave in interesting and dynamic ways. It is our perspective that hierarchical representations are key to task modeling that is supported by a body of existing literature~\citep{bystrom1995task, bystrom2005conceptual, xie2008interactive, xie2009dimensions, jarvelin2015task, cheng2015break, toms2019information, soufan2021untangling}.

%This tree view, albeit grossly simplified, does offer one way to think about task support from a search perspective.

In principle, a search system can offer support at each level of this hierarchy, although in practice search support tends to be small-scale. For example, actions are supported by techniques such as query auto-completion (supporting the current action) or query suggestion (supporting the next action), and these supports are relatively well-studied \cite{cai2016qac}. Some low-level tasks are also supported in search systems: for example, major web search engines offer booking widgets for flights and hotels, directly supporting these small transactional tasks. Mid-level tasks can be supported by, for example, recognizing a flight booking and offering to book a hotel and transport. Although only partially search applications, airline websites routinely offer this. High-level tasks, such as planning an entire vacation, are not at all well supported in software but are routinely supported by (human) agents and delegates.

%\paul{to think about: does this satisfy the desire above, to have an interpretable but general framework? I'm not sure that it does.}

%\chirag{Sections 3.1 and 3.2: expand to incorporate further details/arguments about why `task potato' and operations on it are a great/most reasonable way to work on task-based IR. [Paul, Bhaskar]}

\vspace*{-0.5em}
\subsection{Moves}
The tree of (sub)tasks and actions also suggests certain moves that competent software should make.
To move \emph{left to right} in the tree is to predict or suggest the next thing in a sequence.
To move \emph{up the tree}, action to task or sub-task to super-task, is to recognize a more complex task, having recognized its constituents \cite[e.g.,][]{jiang2014searching,liu2020identifying}. Finally, to move \emph{down the tree} is to decompose a task \cite[e.g.,][]{hoppe1992task,zhang2021learning}. % TO DO more refs here would be nice--paul

For example, by re-ranking search results, \citet{bennett2012modeling} consider short-term and long-term context information for personalization which in our framework corresponds to moving \emph{left to right} for short and longer distances but without explicitly modeling the hierarchy.
Similarly, \citet{mitra2015exploring} considers sessions as paths in query embedding spaces, again moving \emph{left to right} without specifically modeling the hierarchical relationships.
Finally, \citet{sordoni2015hierarchical} use a hierarchical recurrent encoder-decoder architecture to simultaneously model the sequential relationship between terms in a query and between queries in a session.
While they do not consider higher level relationships between search sessions, sub-tasks, and tasks, it may be natural to employ such methods to model task hierarchies.

\vspace*{-1em}
\subsection{Challenges}
\label{sec:challenges}
This model illustrates some challenges we face, if we are to build task-aware search completent in long run. First, some moves around the tree are easier than others. For example, at the time of writing, popular web search engines % I checked Yandex,Yahoo,Naver too--PT
support small, transactional tasks---such as booking a flight---only when the most-recent query looks promising. Research on building longer-term task models is still limited \cite{kotov2011modeling,wang2013learning}, even at the scale of consecutive searches \cite{liao2012evaluating, white2013enhancing}, meaning this move is currently only possible when there is a 1:1 correspondence between task and action (point \circlenum{1} in Figure~\ref{fig:task_potato}).

Some actions are also unobserved, or unobservable, from software, even in practice (\circlenum{2}). For example, a web search engine will most likely be unaware of a searcher's other activity online; all online services will be blind to a face-to-face conversation.

Finally, observed actions are sparse signals and more than one task will have similar steps, so moving up the tree is more difficult than moving down or sideways. We can easily imagine support for decomposing tasks, and can also imagine going across the tree at any level: for example, we could predict the next action given a sequence of actions, or we could predict the next microtask given a sequence of microtasks. It is harder to imagine getting from observed actions to the uppermost (macro-) task or goal~\circlenum{3}, especially when observations are incomplete~\circlenum{2}. We must also note that the data searchers give us will be bound by the affordances we give them; in practice, that means that searchers will express themselves in short keyword phrases (``lhr lax flights'') rather than explain a task (``I need to get to the LA office for Wednesday's big meeting'').

%TO DO ??ISS structures help us predict actions, given a task.??

Challenges for supporting tasks in search therefore include:

\begin{enumerate}[leftmargin=*]
    \item \emph{Representing tasks} in ways that allows the system to take actions. This representation needs to handle tasks at different granularity, with different topics and strategies, and tasks which persist over time.
    
    \item \emph{Observing more task-relevant context}, to better identify and track tasks as they happen. This needs to include tracking across different devices and different timescales, so we can better identify tasks from actions and ``move up'' the tree.

    \item \emph{Developing task-oriented interfaces} that encourage descriptions of task, not need and not short queries; and which support tasks as they happen, either in the search interface or elsewhere. 
\end{enumerate}

%In the next section, we suggest approaches to these problems.
\section{Task modeling}

There are different ways we can extract, represent, and apply task information to address the challenges discussed in Section~\ref{sec:challenges}. In this section, we review some possible approaches we could take in modeling and extracting complex task structures composed of any number of tasks or sub-tasks. %at different granularity from search interactions. %\paul{this sentence is hard to parse: do we mean ``\dots some āpproaches to using search interactions to model and extract complex task structures''?}

\vspace*{-0.5em}
\subsection{Task Representation}
In the model shown in Figure 1, tasks can be defined at different granularity levels. This flexibility provides ways to represent tasks from different theoretical and methodological perspectives. At the same time, it asks for a far-reaching representation capable of modeling work at multiple levels of abstraction~\cite{paterno2004concurtasktrees}. Task descriptions can range from a high level of abstraction to a concrete, granular action-oriented level with precise information need strongly associated to the task. As mentioned by~\citet{paterno1999model}, to build an intelligent task-aware search system, it is necessary to support tasks at each level of the task hierarchy not only from top to down but also from left to right.
%especially in more extended terms \paul{What does ``in more extended terms'' mean?}. 
There are many possibilities to instantiate our task framework by applying diverse supervised and unsupervised techniques depending on the availability of search interaction signals. Assuming that there may be multiple sub-tasks associated with a user's information need and that these sub-tasks could be interleaved across different sessions, a bare tree extraction algorithm has the potential to extract a hierarchical representation of tasks/sub-tasks embedded in search processes as considered by~\citet{mehrotra2015a} (e.g., decomposing a macro task into microtasks as moving down the tree in Figure~\ref{fig:task_potato}). The approach allows us to go across the tree at any level.

Another possible approach could be a vector representation of tasks implicit in search behaviors (i.e., points \circlenum{1} and \circlenum{3} in Figure~\ref{fig:task_potato}% depending on the availability of observable search interactions
) by triangulating observable search events with other situational and contextual information related to the search process. This abstract representation of tasks can especially be helpful in search scenarios where searchers' tasks are not clearly expressed or manifested. For example, existing research has shown how such signals indicate the nature of the task being done \cite[e.g.,][]{choi2016probing, mitsui2019bridging, liu2019task, liu2020identifying}.

To move up and down the task hierarchy, action to task or sub-task to macro-task, it is crucial to know the connections among the contextual components of the search session. Based on the idea that in a real-world information network, proximal nodes in the network structure tend to be similar or related to one another, it is intuitive to visualize user-system interactions initiated by a specific task as a complex graph network structure of users' actions (i.e., query submission, clicks on a document) and systems' reactions (i.e., analyze, retrieve, and display relevant related items). Similarly, queries issued and actions performed by a user and documents viewed within a short time period are more likely to be different stages of the same task, sub-tasks, or sub-sub-tasks; therefore, the search state can be extracted based on similar node representation patterns. Therefore, a sequential heterogeneous graph embedding-based task model \citep[e.g.,][]{fu2020magnn} could potentially capture the structural features of interactive search sessions and represent tasks from observable behavioral signals. This way, the model can represent the macro-task (moving up in the tree) or the next microtask given a sequence of microtasks (moving down or right in the tree).

We have seen several attempts to model search sessions as Markov Decision Proceses~\cite[e.g.,][]{chen2018improving, yang2015query}, Hidden Markov Models~\cite[e.g.,][]{cao2009towards, dungs2017advanced} or Partially Observed Markov Models~\cite[e.g.,][]{yang2018session}. Taking the idea further, we could apply reinforcement learning approaches to learn %left to right in the task hierarchy is
to predict or suggest the next action/task given a sequence of actions or tasks. This is similar to search intent prediction by~\citet{yao2021rlps}. 
%\paul{This is a very long sentence, and I have trouble parsing the second half.}

\vspace*{-0.5em}
\subsection{Inferring Tasks from Observable Events}
Many %of the earlier 
studies used lexical and content-based features, such as the lexical content of queries, for determining topical and task change in the sequence of query formulations. For example,~\citet{verma2014entity} tried to identify entities and clusters of terms related to entities in queries (e.g., using tagging, TF-IDF scoring, term filtering, category terms) to represent a task as a set of terms related to an entity. Other studies have used latent search interaction events to infer tasks (query-based features: query term cosine similarity; URL-based features: URL domain clicked, Jaccard coefficient between clicked URL sets; session-based features: same session and the number of sessions in between, query reformulations, click entropy, query length, post-click actions, and session lengths; temporal features: dwell time for action events). Studies have shown how such signals indicate the nature of the task being performed, even when there is no explicit statement \cite{mitsui2019bridging, liu2019task, liu2020identifying, lucchese2011identifying, wang2013characterizing}. Depending on the availability of search interaction features at a given time, we could exploit several clustering algorithms %and a novel efficient heuristic algorithm 
to extract tasks. %\paul{a ``novel \dots\ algorithm'' suggests we've got something up our sleeve, is that true? Is there something you're hinting at here?}

%\subsection{Interactions and interventions}
\vspace*{-1em}
\section{Applications of task in search}
%\chirag{Change the tone here to talk more about how our proposal and perspective (the way we think about tasks in IR) can help different applications. [Ryen]}

%\chirag{2.5p. for the whole section.}
%\ryen{Contextual search could be another application. Feels like we need to have something in here that is traditional query-based search enhanced with task.}
Task information applications can pave the way for simulating, developing, and evaluating task-aware support. Although existing search systems have improved incredibly and support users with specific factual information tasks, their support is still lacking for complex and exploratory search tasks. Given the nature of these tasks, they need to be decomposed into multiple actionable sub-tasks (i.e., move down the task tree shown in Figure \ref{fig:task_potato}). They may require numerous rounds of interaction (queries/clicks, from a search engine perspective) to complete those tasks \cite{Aula:2010}. Tracking and completing those sub-tasks increases cognitive demands, regardless of user experience level. The task tree can be applied to decompose exploratory and complex tasks into smaller goals, hence reducing cognitive load. This can also help narrow the focus of the assistance offered to the specific task at hand, which could be represented in a semantic space (the so-called ``implicit representations'' referenced earlier) to better identify the task and more fully capture the user's underlying goals and intentions.

In this section, we examine four applications where such considerations of task-based knowledge are valuable.

%During search sessions, users' search interactions have been used to personalized search results; however, the focus is always on query-document matching or developing ranking and re-ranking models to retrieve documents based on searchers' topical interests. However, since queries and search interactions occur in a task context, focusing only on query-document matching or topical-interest matching may be insufficient; rather, it would be more valuable if it is possible to incorporate the task information for more effective search-result ranking and stretch the usability of search systems beyond information extraction to task completion. The abstract task representation mentioned above will be beneficial to model users' search behaviors to generate task models at any level and search situations to find other users attempting similar tasks to complete, identifying documents that appear relevant. Other than traditional query-based search, task representations can be useful in contextual search, conversation search scenarios, and generating proactive support for complex tasks.

%\vspace*{-1em}
\subsection{Contextual Search}
Searches are performed within a situational context. Understanding and modeling this context, especially the current task, is vital for search systems in finding the most relevant information. Task models derived from recent queries and clicks (i.e., the observable actions in the leaf nodes of Figure \ref{fig:task_potato}) within the current session can be applied to improve search engine performance \cite{xiang2010context, shen2005context}. These task representations can assume many forms, including distributions over topical categories \cite{bennett2012modeling} or semantic vectors \cite{mehrotra2017task}.

As we try to model tasks in a short-term search context, we often find ourselves discussing \emph{sessions} (sequences of interactions demarcated by topic or time \cite{jones2008beyond}), which are not exactly the same as tasks (especially given multi-tasking \cite{spink2006multitasking}) but are a reasonable proxy for task in a search setting and are a valuable source of tasks data \cite{liao2012evaluating,liao2014task}. Task models must evolve over time as more evidence is collected about user interests and intentions (implicitly, explicitly, or both) and ideally be transferable across sessions as tasks are suspended and resume over time \cite{agichtein2012b}. %Our task representation from earlier also needs to include these down times between observable actions, both to more fully describe the task and also because they do offer opportunities for additional search assistance (e.g., slow search \cite{teevan2014slow}). 
Other search-related applications of task models that span the leaves of our task tree include personalizing search results \cite{mehrotra2015terms} and generating query suggestions \cite{garigliotti2017generating}.

%\vspace*{-1em}
\subsection{Multi-device Search}
Complex tasks can span both time and space. Another way that the leaves on the task tree can be related is in terms of the devices used. As mentioned in the previous section, there has been some focus in IR on supporting cross-session tasks \cite{agichtein2012b}. Cross-device searching \cite{wang2013characterizing,montanez2014cross}, where people initiate a task at one time and/or on one device and resume it later, perhaps on a different device%(i.e., device A \emph{then} device B)
, is related to cross-session and may be simply because of necessity, but also the device capabilities (e.g., larger display, availability during commute). Supporting both types of searching requires a task representation that is transferable between devices (something more abstract and consistent than a sequence of observable actions). This involves moving up in our task tree, from actions to micro-tasks, sub-tasks, and so on, stopping at the point where the device space can be most fully represented without being so broad that the task representation is meaningless. Multi-device experiences capitalize on the strengths of multiple devices simultaneously to support complex tasks (e.g., recipe preparation, home or auto repair) \cite{white2019multi}. For example, we can combine a smart speaker such as an Amazon Echo with a tablet such as an Apple iPad capitalizes on the far-field speech recognition capabilities of the speaker and the high-resolution display of the tablet. In these experiences, the evolving task representation (implicit, explicit, or both) plays a central role in connecting the devices and providing dynamic context. % during the interaction.
%This particular setup enables people to step through a complex task hands-free via speech, while also having access to the context and supporting information (next steps, media) on the display device. Although some smart speakers now come with displays, there is additional cost involved -- and we can equal, or even exceed, the capabilities of smart displays by combining devices many people already own via, say, a cloud-based orchestrator.

In multi-device scenarios, as with many other task scenarios, task assistance can be offered to users at different stages of the task (e.g., proactively searching for resources related to the current action \cite{nouri2020proactive}) depending on an understanding of the task and the affordances available. This multi-device paradigm can also apply directly to a search context, where, for convenience, people can pose natural language questions to smart speakers via voice, obtain quick answers, and use their smartphones or tablet devices to review supporting information (videos, websites, documents, etc.). For example, a child getting quick responses from a digital assistant (e.g., an answer to a math question) on a smart speaker or smart watch can also be shown explanatory information on a larger display device. %Completing complex search tasks requires that people process and act on complex information, often from different sources and assuming different forms. 
Supporting the use of combinations of devices in multi-device search can provide a way for people to maximize the quality and diversity of the information that they utilize. More fully representing tasks, and their dynamism and context sensitivity, is critical in supporting these multi-device behaviors.

%\vspace*{-1em}
\subsection{Conversational Agents}
%\chirag{Assigned to CHIRAG, BHASKAR}
One of the active areas of application for task-based IR is conversational agents. %Let us envision a scenario with a futuristic intelligent agent.
One can imagine the following conversation happening with an agent over voice %modality
using, for example, a smart speaker or a smartphone. % or some other device that is yet to be invented. %These conversations comprise observational actions but they support the construction of dynamic task representations (going up the task tree) that help better model user intentions and ground future engagement, and the decomposition of tasks into specific action sequences and dialog turns (going down the task tree) to obtain the information the agent requires to offer task assistance.

\begin{quote}
{\bf User}: I think I would like to go do some outside activity today. Do I need to wear a face mask if I go running?\\
{\bf Agent}: It depends where you are running, but if you are concerned about safety or compliance and still want an outdoor activity, may I suggest biking?\\
{\bf User}: Oh.. ya, sure, that could work. Do I need to know anything?\\
{\bf Agent}: While you don't need to wear a mask while biking, you should still bring one with you. There is also a chance of some rain showers, so plan for that. And yes, definitely carry some water.
\end{quote}

Now let us examine what may be going on here. There are four distinct capabilities that we see the agent exhibiting.

\begin{itemize}[leftmargin=*]
    \item {\em Understanding the intention behind a user seeking information.} The agent understands that the user wants to do outdoor activity while being safe.
    %\footnote{Note that this example of was created at the time of the COVID-19 pandemic, when face masks were required for many public activities to reduce the spread of the virus.} 
    This understanding enables the agent to make other recommendations beyond simply answering the question.
    \item {\em Addressing the effects of unknown unknowns (i.e., ``people don't know what they don't know'')}. The user asked ``what do I need to know if I go biking?'', indicating their lack of knowledge about even what may be the right questions to ask. This often happens in human-human interactions. %However, our current systems do not handle such questions well. 
    Here, the agent understands the situation (task), as well as the intention behind that question and responds with relevant suggestions.
    \item {\em Zero-query recommendations}. The user does not ask about weather, but the agent deems it important to convey that information as it may affect the outdoor activity. Also, given the nature of the activity (biking), the agent also recommends carrying water. These are examples of {\em zero-query recommendations}, in which an answer is provided without there being a clear question. Again, doing something like this requires a deep understanding of the situation (task), the user, and their intentions.
    \item {\em Proactive recommendations}. The conversation starts by the user asking a question about running, but rather than completely answering that question, the agent makes a different suggestion (biking), which turns out to be a better one. This is a case of the agent being proactive. In order to go beyond the user's need (at least the expressed need) and provide a relevant and compelling answers or recommendations, an agent needs to be able to understand the purpose behind the potential task, the user's intention behind asking a question, and the world knowledge about how different tasks are executed.
\end{itemize}

In short, to create an intelligent agent like the one envisioned in the scenario above, we need to bring in the following capabilities:

\begin{itemize}[leftmargin=*]
    \item Abstracting out from a query or a question or even an observation to the task and/or context.
    \item Leveraging world knowledge (in this case, public health guidelines and mask mandates).
    \item Generating recommendations from that task/context and weighing whether that would outperform query/question-based recommendation.
    \item Learning how to perform a task.
\end{itemize}

As one can see, much %(not all)
of what we need %to do
revolves around tasks. This is just a simple example of a short conversation. Imagine having discussions (and even debates) about health, politics, and more. Imagine carrying out such conversations across multiple sessions, multiple devices, and multiple people. There are tremendous possibilities here for a giant leap for IR systems. We believe at its core is the notion of task and ways to capture, represent, and address it.

%\vspace*{-1em}
\subsection{Proactive Search and Recommender Systems}
The ability to identify and automatically extract and represent tasks accurately has implications for search or recommender systems in understanding users' information needs at different task levels as well as supporting people in task completion. Therefore, it is crucial to understand how to utilize this knowledge about tasks behind the request to improve a system's offerings to its users. Also, the ability to model users' tasks from their observable actions (at different levels per Figure \ref{fig:task_potato}) unlocks new directions for solving many problems and improving user engagement and satisfaction for building intelligent and proactive systems that can retrieve and recommend information implicitly without requiring explicit queries or other interactions~\cite{dumais2004implicit}. This is important because research has shown that people often struggle to get their tasks done due to a lack of knowledge, motivation, or information literacy~\cite{sarkar2020implicit}. %Besides, as noted in previous sections, search and recommender systems should help people accomplish tasks. Reliably understanding and modeling tasks will enable systems to help people make more progress on their tasks.

The observable actions covered earlier are primarily those taken by the user on their initiative, but this need not always be the case. In mixed-initiative systems, these actions can be prompted by the system or even taken by the system on the user's behalf \cite{horvitz1999principles}, i.e., new leaf actions in the task tree can be proposed or created automatically. The notion of proactive search systems is not new. Letizia \cite{lieberman1995letizia} was one of the earliest applications that provided proactive recommendations during web browsing. Commercially deployed proactive, intelligent systems such as Google Now and Microsoft Cortana can model short-term and long-term search intents and tasks based on search log history~\cite{guha2015user}. In recent times, Song and Guo~\cite{song2016a} proposed proactive recommendations to the user at specific times based on repeated pattern recognition over time. Incorporating task understanding into a proactive system could support users in each task stage and help enable task completion. A task-aware intelligent system could proactively identify potential problems in users' search paths and guide users at various task levels by providing help recommendations or what actions could be executed next to avoid future problems. The aforementioned task representation can be incorporated into various sequence-to-sequence models, probabilistic, or Markov decision-based reinforcement learning models to generate proactive recommendations.
\section{Evaluating Task-Based Applications}
Evaluation is central in IR \cite{kelly2009methods} and this is no different in task-based search and recommendation systems. Many of the same \emph{methodologies} (user studies, simulations, etc.) used in IR to evaluate system performance can be used to evaluate systems to support tasks in search and recommendation settings. Non-task-based IR systems tend to focus on ad hoc retrieval and consider each query independently. Task-based systems consider tasks holistically, spanning multiple queries and/or sessions, the associated context, and task outcomes. The \emph{metrics} used to determine task-based system performance deserve special attention given the focus of these systems on supporting full task processes (not individual queries) and attaining task completion (not only result relevance). We now offer a perspective on methods and metrics for task-based evaluation.%We also need to do more to develop explainable integrated metrics \cite{toms2009isss} that reflect this holistic perspective and consider currently unobservable aspects of task completion and build proxies or (with consent) observe them.

\vspace*{-0.5em}
\subsection{Methodologies}
Many standard evaluation methods (user study protocols, instruments, etc.) apply to the evaluation of task-based systems \cite{kelly2009methods}. %These can be offline, without live users and focused on specific system components in highly controlled settings, or online, with less control but more realism. 
In IR, the Cranfield experiments \cite{cleverdon1967cranfield} and TREC \cite{voorhees2005trec} have driven considerable progress, including in tasks research \cite{yilmaz2015overview}. %These initiatives abstract away many human elements and focus on top resources and the document to query match. 
Beyond Cranfield and TREC, evaluation in IR must now take a broader view on tasks, users, and context \cite{jones2006s}, to improve experimental realism and the reliability of conclusions drawn. %Many of the methods traditionally used to evaluate IR systems (user studies, simulations, search log analysis, flighting, etc.) also apply in a task-based setting. 
Methods such as living laboratories \cite{kelly2009evaluation} bridge user- and system-centered research via resources, tools, and infrastructure for collaborative experimentation \cite{balog2014head}. Mixed methods studies can provide a more complete picture of task performance, albeit with more complexity and greater cost than single-method studies. As mentioned earlier, tasks can extend over time and be part of larger macrotasks. This additional context should also factor into task-based evaluation \cite{dumais2009evaluating}.

\vspace*{-0.5em}
\subsection{Metrics}
\label{sec:metrics}
Evaluating systems on the basis of search task performance has been explored for decades \cite{hersh1996task}. %To perform task-based evaluation, we need to consider both process metrics and outcome metrics. We may also consider pre-task metrics, associated with preparation and preparedness, which are more user- and scenario-dependent and not a focus here. 
All metrics make assumptions about task behavior, which must be validated \cite{dupret2008user}. Conceptualizing tasks and creating task models are important in determining appropriate task-based evaluation metrics. It is insufficient to solely target system functionality (or even more narrowly: specific components) when systems and users must collaborate to complete tasks successfully \cite{bates1990should}. We should evaluate task-based systems \emph{holistically} to reach actionable conclusions and understand system performance \cite{balog2015task}. We discuss that now, targeting task processes and task outcomes. %We provide examples of process and outcome metrics, focused initially what we believe are important starting points to evaluate task-based systems in a context of use.

%\vspace*{-0.5em}
\subsubsection{Task Processes}
Process metrics are focused on how people attempt to complete the task, regardless of the task outcome. They include: (1) \emph{Task completion time}, both actual time and perceived time. Time has been used in search evaluation \cite{xu2009evaluating,fox2005evaluating}. Task has been shown to affect document dwell times \cite{kelly2004display,white2006study}. \citet{smucker2012time} studied time from the perspective of gain per unit time. Perceived time can differ from stopwatch time per factors such as attentional demand \cite{csikszentmihalyi1990flow}; %Prospectively, methods such as GOMS \cite{card2018psychology} provide quantitative and qualitative predictions of how people will use a system, including timing \cite{john1996goms}; 
(2) \emph{Effort expended} to complete the task (e.g., the number of actions taken, recommendations reviewed, dialog turns). In search, effort typically describes the number of searches or clicks \cite{azzopardi2013query,cooper1968expected}. \citet{kelly2015effort} discussed the relationship between expected and experienced effort (e.g., if experienced effort is less than expected, the task is considered easy). Effort underlies many user models in IR evaluation \cite[e.g.,][]{jarvelin2008discounted,moffat2008rank}. \citet{kiseleva2016understanding} showed that user satisfaction is negatively correlated with the amount of effort to complete a task: more effort means less user satisfaction; (3) \emph{Engagement} covers the connection between the user and the system, spanning emotional, cognitive, and behavioral aspects \cite{jacques1996nature}. It is affected by many factors, including user and task characteristics, user experience, and biases \cite{o2008user}. It can be a goal in task-based systems (e.g., in open-domain dialog \cite{huang2020challenges}) but also a side effect (e.g., in task-oriented dialog systems \cite{chen2017survey}), and; (4) \emph{Progress} through the task. %where tasks can be self-contained (simple, atomic tasks performed within a single session) or they can span sessions or devices (complicating progress monitoring).
Detecting task completion can be straightforward for some tasks, e.g., transactional tasks, but complex for others, e.g., learning tasks \cite{white2010assessing}. Progress can be tracked using dedicated tools \cite{bellotti2004studies} or inferred \cite{white2019task_2}. Recent research has built benchmarks for measuring task progress in digital assistants \cite{liono2019building}. Task-oriented dialog systems, focus on metrics such as number of slots filled ($x$ of $y$) \cite{budzianowski2018multiwoz}. %; $y$ is known in stepwise tasks, such as cooking, reservations, and so on.
These four popular metrics are broadly applicable, are easy to define in task-based search and recommendation settings, and can be computed at low-cost at large scale. There are other metrics including cognitive load \cite{beaulieu1997experiments}, learning \cite{rieh2016towards}, affect \cite{feild2010predicting}, and usability \cite{albert2013measuring}, which are more challenging to define and measure. %All are important and necessary to develop a comprehensive picture of task performance.

\vspace*{-1.5em}
\subsubsection{Task Outcomes}
Outcome metrics focus on the product of tasks, either a real outcome (e.g., task completion) or a user-perceived outcome (e.g., satisfaction). Salient examples include: (1) \emph{Task utility}, denoting the value of information obtained to complete the task, e.g., relevance \cite{mizzaro1997relevance}. Relevance is affected by task stage \cite{taylor2007relationships} and relevance metrics help estimate support for task completion \cite{moffat2017incorporating}. Relevance metrics are usually computed per query but session-level metrics must also be considered in task scenarios \cite{luo2013water}, as must task support beyond result pages \cite{downey2008understanding}. Relevance is personal and situational \cite{saracevic2007relevance} and task-based evaluation must consider that, e.g., during contextual search \cite{bennett2012modeling}; (2) \emph{Satisfaction} with the outcome of the task and the process, often modeled at the task/session level \cite{hassan2011task}. Satisfaction is non-binary and impacted by task and user effects \cite{kelly2004display, kim2014modeling, white2006study} and even query position in the session \cite{jiang2016correlation}. More observations of on-task behavior enable more accurate models of satisfaction \cite{huang2012improving,kiseleva2016understanding}, and; (3) \emph{Task success}, covering whether task objectives were accomplished. This relates to satisfaction but not entirely and can be modeled based on behavioral signals \cite{hassan2010beyond}. Completion events such as in-world activities may be unobservable to online systems, making it difficult to measure task success, although proxies e.g., conversions \cite{brovman2016optimizing} may offer insight. Other task outcome metrics, including novelty and diversity \cite{clarke2008novelty}, creativity \cite{shneiderman2000creating}, and adoption and retention, e.g., search engine switching \cite{white2010modeling} and sustained use over time \cite{dumais2013task}, are promising but are also less well defined and require data that can be difficult to obtain.

%\vspace*{-1em}
%\vspace*{-1em}
\subsection{Additional Considerations}
There are many other metrics that can apply to task-based systems including robustness, privacy, adaptivity, and scalability \cite{shani2011evaluating}. In developing task-based metrics, we also must consider user models (e.g., personas) and task models (e.g., search strategies and goals). Task performance is affected by many factors, including intrinsic properties of the task (e.g., nature of the task \cite{Mitsui2018}, topic \cite{mehrotra2017hey}, difficulty \cite{wildemuth2014studies}, complexity \cite{bystrom1995task}) as well as extrinsic properties such as user attributes (e.g., expertise \cite{white2009characterizing}, familiarity \cite{kelly2002effects}), the situation \cite{jarvelin2015task,ingwersen2006turn,shah2017social}, and other factors such as meta-cognitive skills in task planning and reflective assessment \cite{biggs1988role}. We must also understand the nature of the user experience, which impacts how metrics are defined and interpreted. Metrics also interact, e.g., effort affects satisfaction \cite{yilmaz2014relevance} and they trade off, e.g., time taken versus coverage \cite{toms2009isss}. Metrics must be contextualized, e.g., not all effort is detrimental and more effort could also mean more learning. %more about different perspectives, enabling better decision making.
%Task activity may extend beyond engagement with information systems and ways to collect that data (with user consent) is important.

Task support systems also contain multiple connected components \cite{myers2007intelligent}. Evaluating per component performance has limited value in appraising what the user would experience \cite{toms2009isss}; hence our focus here on holistic metrics. However, the metrics may not be correlated \cite{frokjaer2000measuring}. Integrated metrics combine multiple variables \cite{tague1992measuring,toms2005searching,o2008user}, although these can be difficult to interpret. %Although powerful, the intuition behind these metrics is unclear, as is the interpretation of the resulting models; necessitating follow-up with human subjects. 
Sets of metrics are commonly employed in the evaluation of task-oriented dialog systems \cite{walker1997paradise} and defining such a set of metrics that are agreed upon by the community could help evaluate task-based search and recommendation systems. Meta-analysis frameworks \cite{sakai2012evaluation,amigo2018axiomatic} analyze the extent to which metrics capture key properties and align with user preferences; they may also be applicable here.
%\vspace*{-1em}
\section{Task Futures}
%\chirag{Give a couple of examples of what comes next and how that work ties to what we proposed here. [Chirag]}

Considering user tasks in IR is not a new idea, but every new generation of IR students and scholars seem to encounter it in a new light -- sometimes leading to groundbreaking advancements, and other times redoing or incrementally adding to previous work. With the increasing attention to and importance of emerging IR applications, we believe the time is ripe for a new generation of scholars to not only rediscover task-based IR, but also take a conceptual and practical leap to finally realize the vision of supporting users in accomplishing their tasks, regardless of their information literacy or specificity in queries. We now consider some future directions and conclude by discussing key ethical considerations.%In the next subsection we highlight some of the larger research directions and threads that we see possible as the next steps in this area. We conclude the perspective paper with a discussion on ethical considerations.

%\vspace*{-1em}
\subsection{Research Threads and Directions}
Here, we identify some big challenges, each suitable for one or more PhD dissertations or grant proposals:

%\nick{Is there any way to embed, or relate, this list to the framework and the other proposals made up to this point?}
\begin{itemize}[leftmargin=*]
\item Task understanding
    \begin{itemize}[leftmargin=*]
        \item Formalize and validate various task representations (both implicit and explicit, as mentioned earlier), potentially tying them to different contexts or applications.
        \item Investigate different ways to use contextual information (e.g., spatiotemporal signals, concurrent running applications) to better understand tasks.
        \item Extend task understanding across multiple sessions and/or devices.
        \item Attributing and aggregating observed actions into higher-level tasks (moving up the task tree).
    \end{itemize}
\item Task support
    \begin{itemize}[leftmargin=*]
        \item Make task a first class object in search support, e.g., surface guided tours in response to exploratory queries.
        \item Provide support for task completion (not just providing search results), including recommending search as a means of task completion, where appropriate.
        \item Integrate IR applications with existing task applications such as Microsoft To Do and Google Tasks, as well as email and calendar, to seamlessly surface task-related information and actions.
        \item Better support complex tasks comprising multiple steps, including decomposing complex tasks into more manageable sub-tasks, and supporting search across multiple sessions and/or devices.
        \item Support team tasks (direct collaboration, sub-task assignment, load balancing, etc.) in addition to individual tasks.
        \item Cooperate with users directly, e.g., task-oriented dialog systems, to address tasks more explicitly and also to better educate users about the role of IR systems in solving tasks.
        \item Explore task automation, starting with frequent or recurring tasks, e.g., travel planning, finding job opportunities, and researching a socio-political issue, including extending work on standing queries \cite{morris2007s} and slow search \cite{teevan2014slow}.
    \end{itemize}
\item Task data and experimentation
    \begin{itemize}[leftmargin=*]
        \item Provide lightweight task capture mechanisms, as ground truth for machine learning models and to build trust in task assistance with users by giving them agency over what task-related information is shared with the system.
        \item Find ways to uncover more unobservable events related to the task process (triangulate data sources, with user consent).
        \item Create shared datasets and challenges, with user consent, to promote task-related research and mitigate risk of leaking sensitive data via methods such as differential privacy \cite{dwork2008differential}.
    \end{itemize}
\end{itemize}

We believe the framing device presented in this paper (Figure \ref{fig:task_potato}) as well as our proposals for how such a device can be useful in modeling and using task in search applications (Sections 4 and 5) can help for at least some of these directions. For example, the task tree structure along with the formulations of various moves presented in Section 3 can be used to define a set of support actions (e.g., offer within-task query recommendations with traversal to a sibling node, suggest related tasks with a jump to a new parallel branch in the tree) in interactive search. This structure can be comprised of (1) identifying which part of this task tree the user is at a given moment; (2) deciding what could be the next set of sub/super/related tasks could be from this tree; and (3) making and revising recommendations based on user actions (moves).

%As a final thought, we once again note the importance of ethical considerations that need to go into each of these tasks, specifically ensuring that we use task understanding and support as a way to bring more equity and diversity and not reduce them.

%\ryen{This discusses where the field should go. This should include things like (in no particular order): 

%\begin{itemize}
%\item validating task representations 
%\item provide support for task completion (not just providing search results), including recommending search as a means of completion
%\item integration with current task applications (to-do, email, etc)
%\item use context to better understand tasks
%\item better support complex tasks comprising multiple steps (including decomposing complex tasks into sub-tasks)
%\item consider tasks that span time and/or devices (including multi-device experiences)
%\item provide lightweight task capture mechanisms, as ground truth for ML and to build trust with users
%\item cooperate with users directly, e.g., task-oriented dialog systems
%\item make task a first class object is search, e.g., surface guided tours in response to exploratory queries
%\item support team tasks (direct collaboration, sub-task assignment, sub-task sequencing, load balancing, etc.) in addition to individual tasks
%\item create shared datasets, challenges, living labs, etc. to boost task research
%\item explore task automation, starting with frequent or recurring tasks
%\end{itemize}
%}

\vspace*{-0.5em}
\subsection{Ethical Considerations}
%\chirag{We have to be careful with task derivation and recommendations based on that derivation. We may be influencing the user too much and inadvertently end up making search too homogeneous. Assigned to CHIRAG, ...}
%While the previous subsection provided several ideas for capturing and representing tasks, it assumed that doing so is only in a user's interest. 
Capturing and representing tasks can have benefits, but at what cost?
Many scholars have argued that low information literacy can lead to users not being able to fully utilize the available information or the tools to their most potential \cite{russell2019joy}. Even for users with reasonable or high information literacy, they often ``don't know what they don't know'' \cite{belkin1980anomalous, shah2018information}. In other words, if an IR system is relying on a user explicitly and at least partially expressing their information needs in order to provide them results or recommendations, it is likely to face challenges serving these populations of users. Extracting and using task information, and being proactive in search can help such users \cite{white2006study}. However, what is often ignored are the ethical considerations and responsibility of researchers and developers. 

As we move toward systems that go beyond serving explicit requests from users, with task-based IR systems %as discussed here
being one of the examples, there are dangers in how such systems could unduly influence user behaviors and nudge them in ways that perpetuate bias and a false sense of trust. With rapid development in artificial intelligence techniques that are being deployed in search systems, those systems become less and less \textit{trustworthy}, even while usually remaining \textit{trusted}~\cite{pan2007google}. The feedback loops created between systems recommending information and users selecting among recommendations make the selections less and less useful for training: we are no longer observing human behavior, but controlling it~\cite{lutz2019digital,neri2020role}. This effect, along with other systemic effects, means that the datasets on which models are trained include significant biases~\cite{tommasi2017deeper,geirhos2018imagenet,torralba2011unbiased}.

This vicious cycle of a system getting ahead of user requests to recommend results and the users clicking on them as they either lack motivation or enough information literacy can be manifested in several ways. For instance, this proactive, task-based recommendation could lead to a search engine promoting its own services and tools simply because it has access to a lot more data and insights about those entities than those from their competitors.

Identifying and modeling tasks may call for more data collection from more people, even those who do not actively use the system. We need to balance the need for more data and the dangers of ubiquitous data collection such as surveillance capitalism and other forms of abuse~\citep{harwell2019doorbell,haskelldowland2021airtag,zuboff2019surveillance}.

As task modeling inherently necessitates predicting users' next actions/needs, we must consider the cost of false prediction (e.g., requiring user to perform even more actions to counter the system's false beliefs regarding user goals or intentions). A related question is how to recognize and respect user agency in their tasks and not overtly influence their course of action.

We should also not assume that a task modeling system can easily identify and address a singular objective or interest. When different stakeholder interests are involved, how do we balance across the different dimensions and control for unintended consequences? For example, a tool that makes it really easy to book a flight may unintentionally discourage users to do more research that may lead to cheaper tickets. Finally, if task modeling is inherently complex and resource intensive, it might mean that system designers need to prioritize which tasks they support, raising questions about fairness across different user populations. In short, explicating and using task information, while important and desired, must be done with ethical issues in mind. We should, in general, create a practice of integrating such considerations from the outset rather than trying to address them later or fix problems resulting from not considering them as a posthoc activity.

\begin{acks}
This work was partially supported by National Science Foundation (NSF) grant III-1717488.
\end{acks}

\balance

%\newpage

\bibliographystyle{ACM-Reference-Format}
\bibliography{references}
\balance

\end{document}